\documentclass[thmsa,a4paper,final,notitlepage,12pt,onecolumn]{article}
\usepackage{amssymb}

\usepackage{sw20lart}

\newlength{\defaultparindent}
\input{tcilatex}
\begin{document}

\title{\textbf{On The Significance of Bell's Locality Condition}}
\author{A. Shafiee\thanks{%
E-mail: shafiee@theory.ipm.ac.ir} \ and M. Golshani\thanks{%
E-mail: golshani@ihcs.ac.ir} \\
Institute for studies in Theoretical Physics and Mathematics \\
( P. O. Box 19395-5531, Tehran, Iran )}
\maketitle

\begin{abstract}
\textbf{Reviewing the general representation of a stochastic local hidden
variables theory in the context of an ideal Bohm's version of the EPR
experiment, we show explicitly that the violation of Bell's locality
condition is due to the assumption of ``outcome independence'' at the hidden
variables level. Also, we show that if we introduce determinism, the
assumption of outcome independence will be allowed.}
\end{abstract}

\section{Introduction}

Since the derivation of original Bell's inequality [1], this inequality
(especially in the form of CH [2] \& CHSH [3] inequalities) has been derived
in various ways. It has been extended to more than two particles and to the
cases where the initial state's spin is higher than zero. But what is
generally accepted as an interpretation of all forms of Bell's inequality is
that there exists an incompatibility between any local hidden variables
theory with quantum mechanical predictions, and this is due to non-locality.

Because of the central role of the locality condition in Bell's theorem,
there has been some efforts to clarify its meaning by considering other
conceptual possibilities. One of the most influential works done in this
context is due to Jarrett who obtained Bell's locality condition by the
conjunction of two independent conditions [4]. Shimony called the first
condition ``outcome independence`` [5], which states that in the Bohmian
version [6] of the EPR experiment [7] (hereafter called EPRB), and for
definite settings on both sides, the probability of getting a result on one
side is independent of the result on the other side. The second condition
called ``parameter independence`` by Shimony, holds that the result of a
given measurement is statistically independent of the setting of the distant
measuring apparatus.

We shall discuss these points in more detail in this paper, but what is
important here is that the violation of Bell's inequality implies that
either parameter or outcome independence or both must fail , but it can not
tell us which of the two conditions is violated. So far, no generally
accepted solution has been found for this problem [8].

In section 2, we show explicitly that showing the representation of the
stochastic local hidden variable theories as done by CH [2] and Bell \textit{%
et al}. [9], there is no stochastic outcome independent hidden variables
theory which can reproduce all the predictions of quantum mechanics. To show
the inconsistency, we restrict ourselves to an ideal EPRB thought experiment
and define a condition applicable at the quantum level to prove a
fundamental relation (relation (19)) in the context of a hidden variables
theory, without using any inequality. Then, in section 3, we show that the
outcome independence condition can only be considered in the deterministic
hidden variables theories.

\section{Stochastic local hidden variables theories}

Let us consider a system consisting of a source, which emits two spin 1/2
particles. The spin of particles 1 and 2 are measured in the directions $%
\widehat{a}$ and $\widehat{b}$, respectively. Ignoring the possibilities
concerning the missed detections (as is the case for an ideal thought
experiment), one can use quantum mechanics to calculate various
probabilities. For example, the joint probability for the spin component of
particle 1 in the direction $\widehat{a}$ being $r$ ($r=$ $\pm $ $1$ in
units of $\dfrac{\hslash }{2}$ ) and the spin component of particle 2 in the
direction $\widehat{b}$ being $q$ ($q=$ $\pm 1$ in units of $\dfrac{\hslash 
}{2}$) is equal to

\begin{equation}
P^{(12)}(\sigma _{a}=r,\sigma _{b}=q|\widehat{a},\widehat{b},\Psi _{0})=%
\frac{1}{4}(1-rq\cos \theta _{ab})
\end{equation}
where $\Psi _{0}$ is the singlet state representing the initial state wave
function and $\theta _{ab}$ is the angle between $\widehat{a}$ and $\widehat{%
b}$ .

Noticing the relation between the average values and the probability
measures, one can write the joint probability defined in (1) as [10], 
\begin{equation}
P^{(12)}(\sigma _{a}=r,\sigma _{b}=q|\widehat{a},\widehat{b},\Psi _{0})=%
\frac{1}{4}\left[ 1+r<\sigma _{a}>+q<\sigma _{b}>+rq<\sigma _{a}\sigma
_{b}>\right]
\end{equation}
Here, $<\sigma _{a}>$ is the expectation value of the spin component of
particle 1 along $\widehat{a}$ , $<\sigma _{a}\sigma _{b}>$ is the
expectation value of the product of the values of the spin components of
particles 1 and 2 along $\widehat{a}$ and $\widehat{b}$, respectively, and $%
<\sigma _{b}>$ is the expectation value of the spin component of particle 2
along $\widehat{b}$.

Summing over $q$ or $r$ in (2), yields marginal probability measures for
particle 1 or 2, respectively,

\begin{equation}
P^{(1)}(\sigma _{a}=r|\widehat{a},\Psi _{0})=\frac{1}{2}\left[ 1+r<\sigma
_{a}>\right]
\end{equation}
and

\begin{equation}
P^{(2)}(\sigma _{b}=q|\widehat{b},\Psi _{0})=\frac{1}{2}\left[ 1+q<\sigma
_{b}>\right]
\end{equation}

To have a realistic interpretation of quantum mechanics we seek a stochastic
local hidden variables theory [2, 9] in which $\lambda $ represents a
collection of hidden variables which determine the complete state of the
particles and which belongs to a space $\Lambda $ on which the probability
measures can be defined. Here, one can assume that a spin component of a
particle has a definite value even before it is measured, and that the
statistical results of the spin measurements of one particle do not depend
on the values of the spin components of the other particle. Thus, if the
spin component of particle 1 is measured first, we have

\begin{equation}
p^{(2)}(\sigma _{b}=q|\widehat{a},\widehat{b},\sigma _{a}=r,\lambda
)=p^{(2)}(\sigma _{b}=q|\widehat{a},\widehat{b},\lambda )
\end{equation}
where $p$ is the symbol of the probability measures on space $\Lambda $.
This relation is equivalent to outcome independence. According to (5), the
probability for particle 2 having spin component along $\widehat{b}$ equal
to $q$ is independent of the value of the spin component of particle 1 along 
$\widehat{a}$, if the complete state $\lambda $ and the distant settings are
given. Furthermore, if one assumes Shimony's parameter independence, one gets

\[
p^{(12)}(\sigma _{a}=r,\sigma _{b}=q|\widehat{a},\widehat{b},\lambda
)=p^{(1)}(\sigma _{a}=r|\widehat{a},\lambda )\,p^{(2)}(\sigma _{b}=q|%
\widehat{b},\lambda ) 
\]
\begin{equation}
=\frac{1}{4}\left[ 1+rE^{(1)}(\widehat{a},\lambda )+qE^{(2)}(\widehat{b}%
,\lambda )+rqE^{(12)}(\widehat{a},\widehat{b},\lambda )\right]
\end{equation}
where,

\begin{equation}
p^{(1)}(\sigma _{a}=r|\widehat{a},\lambda )=\frac{1}{2}\left[ 1+rE^{(1)}(%
\widehat{a},\lambda )\right]
\end{equation}
and

\begin{equation}
p^{(2)}(\sigma _{b}=q|\widehat{b},\lambda )=\frac{1}{2}\left[ 1+qE^{(2)}(%
\widehat{b},\lambda )\right]
\end{equation}

Here, $E^{(1)}(\widehat{a},\lambda )$ and $E^{(2)}(\widehat{b},\lambda )$
are, respectively, the average values of the spin components of particle 1
along $\widehat{a}$ and particle 2 along $\widehat{b}$ , and $E^{(12)}(%
\widehat{a},\widehat{b},\lambda )=E^{(1)}(\widehat{a},\lambda )E^{(2)}(%
\widehat{b},\lambda )$ represents the average value of the product of the
spin components of particles 1 and 2 along $\widehat{a}$ and $\widehat{b}$,
respectively. The relation (6) is known as Bell's locality condition.

Multiplying $E^{(12)}(\widehat{a},\widehat{b},\lambda )$ through the
probability density $\rho (\lambda )$ and integrating over $\lambda $ ($%
\int_{\Lambda }\rho (\lambda )\ d\lambda =1$ ), we get the corresponding
expectation value at the quantum level, i.e.,

\begin{equation}
<\sigma _{a}\sigma _{b}>=\int_{\Lambda }E^{(12)}(\widehat{a},\widehat{b}%
,\lambda )\,\rho (\lambda )\ d\lambda
\end{equation}
Similar relations hold for $\langle \sigma _{a}\rangle $ and $\langle \sigma
_{b}\rangle $, regarding $E^{(1)}(\widehat{a},\lambda )$ and $E^{(2)}(%
\widehat{b},\lambda )$, respectively.

As a consequence of the consistency conditions in the laws of probability,
we also have for the conditional probability of the second particle [11],

\begin{equation}
P^{(2)}(\sigma _{b}=q|\widehat{a},\widehat{b},\sigma _{a}=r,\Psi _{0})=\frac{%
\int_{\Lambda }p^{(1)}(\sigma _{a}=r|\widehat{a},\lambda )\,p^{(2)}(\sigma
_{b}=q|\widehat{b},\lambda )\,\rho (\lambda )\ d\lambda }{P^{(1)}(\sigma
_{a}=r|\widehat{a},\Psi _{0})}
\end{equation}

What we want to show explicitly now is that the foregoing framework for the
representation of a stochastic local hidden variables theory is not
successful, even if we accept that superluminal influences do not exist.

To clarify this matter, we look at some of the fundamental requirements
which are relevant to Bell-type ideal experiments. First, we consider two
dynamical variables $S_{1}$ and $S_{2}$ which take values $s_{1}$, $s_{2}$ = 
$\pm $ 1. Then, we define the exchangeability condition on $s_{1}$ and $%
s_{2} $ for the conditional probability measures as [12],

\begin{equation}
\mathsf{P}(S_{2}=s_{2}|S_{1}=s_{1})=\mathsf{P}(S_{2}=s_{1}|S_{1}=s_{2})
\end{equation}
Now, one can show that by imposing this condition, we have the following
result (see appendix):

\begin{equation}
\mathsf{P}(S_{2}=s_{2}|S_{1}=s_{1})=\frac{1}{2}\left[
1+s_{1}s_{2}<S_{1}S_{2}>\right]
\end{equation}

Under the conditions assumed in the EPRB problem (i.e., $\langle \sigma
_{a}\rangle =\langle \sigma _{b}\rangle =0$) if we define $S_{1}\equiv
\sigma _{a}$ and $S_{2}\equiv \sigma _{b}$ (when one defines $S_{1}\equiv
\sigma _{b}$ and $S_{2}\equiv \sigma _{a}$, the argument remains valid for
the probability measure of the first particle), the relation (12) takes the
following form

\begin{equation}
P^{(2)}(\sigma _{b}=q|\widehat{a},\widehat{b},\sigma _{a}=r,\Psi _{0})=\frac{%
1}{2}\left[ 1+rq<\sigma _{a}\sigma _{b}>\right]
\end{equation}
where $<\sigma _{a}\sigma _{b}>=-\cos \theta _{ab}$.

According to (9), $<\sigma _{a}\sigma _{b}>$ has a clear physical
interpretation in a stochastic (outcome independent or dependent) hidden
variables theory and $r$ as well as $q$ are completely defined in this
framework. Thus, the right hand side of (13) is completely defined in a
stochastic hidden variables theory and can be represented as a new relation,

\begin{equation}
f(r,q|\widehat{a},\widehat{b},\lambda )=\frac{1}{2}\left[ 1+rq\,E^{(12)}(%
\widehat{a},\widehat{b},\lambda )\right]
\end{equation}
where $f(r,q|\widehat{a},\widehat{b},\lambda )$ satisfies the following
relation,

\begin{equation}
P^{(2)}(\sigma _{b}=q|\widehat{a},\widehat{b},\sigma _{a}=r,\Psi
_{0})=\int_{\Lambda }f(r,q|\widehat{a},\widehat{b},\lambda )\ \rho (\lambda
)\ d\lambda
\end{equation}

Using the definition of $E^{(12)}(\widehat{a},\widehat{b},\lambda )$ as

\[
E^{(12)}(\widehat{a},\widehat{b},\lambda )=\stackunder{r,q=\pm 1}{\sum }%
rq\,p^{(12)}(\sigma _{a}=r,\sigma _{b}=q|\widehat{a},\widehat{b},\lambda ) 
\]
and substituting it into (14), one obtains 
\begin{equation}
f(r,q|\widehat{a},\widehat{b},\lambda )=p^{(12)}(\sigma _{a}=r,\sigma _{b}=q|%
\widehat{a},\widehat{b},\lambda )+p^{(12)}(\sigma _{a}=-r,\sigma _{b}=-q|%
\widehat{a},\widehat{b},\lambda )
\end{equation}
This shows that $f(r,q|\widehat{a},\widehat{b},\lambda )$ is a probability
measure on the space $\Lambda $. One can also deduce that

\begin{equation}
f(r,q|\widehat{a},\widehat{b},\lambda )+f(r,-q|\widehat{a},\widehat{b}%
,\lambda )=1
\end{equation}
Relation (17) shows that $f(r,q|\widehat{a},\widehat{b},\lambda )$ is a
two-variable probability measure, so that if we sum over one variable and
fix the other, the result should lead to unity.

There is a subtle point that necessitates using an additional assumption in
the relation (16). Let us consider, e.g., the functions $f(+1,+1|\widehat{a},%
\widehat{b},\lambda )$ and $f(+1,-1|\widehat{a},\widehat{b},\lambda )$.
According to the relation (17), these functions are the probability measures
whose values are defined in a set of the ideal experiments in which the
overall number of detections should contain all the results $(+1,+1)$, $%
(+1,-1)$, $(-1,+1)$ and $(-1,-1)$ for $r$ and $q$, respectively. If the
overall number of detections reduces to the first two results $(+1,+1)$ and $%
(+1,-1)$, the functions $f(+1,+1|\widehat{a},\widehat{b},\lambda )$ and $%
f(+1,-1|\widehat{a},\widehat{b},\lambda )$ cannot be specified in principle.
This, in itself, makes no problem. The problem arises when one takes the
relation (15) into account. When only the two-fold detections $(+1,+1)$ and $%
(+1,-1)$ are considered, the left hand side of the relation (15) is
completely defined for $q=+1$ (or $-1$) and $r=+1$. This shows that one
should be able to determine the function $f(r,q|\widehat{a},\widehat{b}%
,\lambda )$ for the two-fold detections too, when the variables $r$ and $q$
are specified. The only kind of a two-variable probability measure $f(r,q|%
\widehat{a},\widehat{b},\lambda )$ which satisfies both the relations (15)
(which necessitates that $f$ should be defined in a two-fold probability
space) and (17) (which is the normalization condition for $f(r,q|\widehat{a},%
\widehat{b},\lambda )$) is a conditional one. Thus, we conclude that

\begin{equation}
f(r,q|\widehat{a},\widehat{b},\lambda )=p^{(2)}(\sigma _{b}=q|\widehat{a},%
\widehat{b},\sigma _{a}=r,\lambda )
\end{equation}
Replacing (18) in (14), we get

\begin{equation}
p^{(2)}(\sigma _{b}=q|\widehat{a},\widehat{b},\sigma _{a}=r,\lambda )=\frac{1%
}{2}\left[ 1+rq\,E^{(12)}(\widehat{a},\widehat{b},\lambda )\right]
\end{equation}
This shows that the relation (5) cannot hold if the predictions of a
stochastic hidden variables theory are to coincide with the results of the
standard quantum mechanics, as is the case in relation (15). Consequently,
we are confronted with the breakdown of outcome independence condition.

The relation (19) implies that $E^{(2)}(\widehat{a},\widehat{b},\lambda
)=E^{(1)}(\widehat{a},\lambda )E^{(12)}(\widehat{a},\widehat{b},\lambda )$
(see appendix). One can also show that $E^{(1)}(\widehat{a},\widehat{b}%
,\lambda )=E^{(2)}(\widehat{a},\lambda )E^{(12)}(\widehat{a},\widehat{b}%
,\lambda )$, if we choose $S_{1}\equiv \sigma _{b}$ and $S_{2}\equiv \sigma
_{a}$, in relation (12). In an ideal EPRB experiment, however, the mean
values $E^{(1)}$, $E^{(2)}$ and $E^{(12)}$ should not depend on our decision
about which particle's spin is measured first. Thus, one can conclude that $%
E^{(1)}=$ $E^{(2)}=0$ for every $\widehat{a}$, $\widehat{b}$ and $\lambda $ (%
$E^{(12)}$ cannot be taken to be always zero, because of the relation (9)).

The relation (19) depends on $\widehat{a}$, and this is an indication of a
kind of influence of the experimental arrangement for the measurement of the
spin component of particle 1 along $\widehat{a}$, on the particle 2. Such an
influence is only meaningful on the basis of the violation of outcome
independence. Shimony has argued that the foregoing violation would not
necessarily allow superluminal signals to be sent, and there remains a
possibility of subluminal signaling in an outcome dependent stochastic
hidden variables theory [5]. The inference that Bell's theorem does not lead
to the superluminal signalling has been analyzed by some people before
[13-16]. Here, we have reached the same conclusion by a completely different
approach.

\section{Deterministic hidden variables theories}

There exists an exceptional situation, for which the exchangeability
condition cannot be inferred from (19). This possibility occurs when $r$ and 
$q$ are constrained by some deterministic conditions.

Assuming determinism in an outcome independent hidden variables theory (as
is the case in Bohm's theory), the knowledge of $\lambda $ determines the
values of each particle's spin along any direction. Thus, the probability
measures are either zero or one, and (regardless of considering parameter
independence) one can use definite values $A(\widehat{a},\widehat{b},\lambda
)$ and $B(\widehat{a},\widehat{b},\lambda )$ instead of the average values $%
E^{(1)}(\widehat{a},\widehat{b},\lambda )$ and $E^{(2)}(\widehat{a},\widehat{%
b},\lambda )$, respectively. Then, if the spin components of both particles
take the values $\pm 1$ in a deterministic way, we have

\begin{equation}
r\,A(\widehat{a},\widehat{b},\lambda )=+1\quad and\quad q\,B(\widehat{a},%
\widehat{b},\lambda )=+1\ \ for\ every\ \widehat{a},\,\widehat{b}\ and\
\lambda .
\end{equation}
In this situation, one can deduce that $p_{DET}^{(1)}(\sigma _{a}=r|\widehat{%
a},\widehat{b},\lambda )$ is equal to one for $r\,A(\widehat{a},\widehat{b}%
,\lambda )=+1$, and zero for $r\,A(\widehat{a},\widehat{b},\lambda )=-1$,
which means the last case never occurs. A similar result holds for $%
p_{DET}^{(2)}(\sigma _{b}=q|\widehat{a},\widehat{b},\lambda )$. Here, $r$
and $q$ are constrained by $A(\widehat{a},\widehat{b},\lambda )$ and $B(%
\widehat{a},\widehat{b},\lambda )$, respectively. Consequently, the relation
(19) should be replaced by the following relation,

\[
p_{DET}^{(2)}(\sigma _{b}=q|\widehat{a},\widehat{b},\sigma _{a}=r,\lambda )=%
\frac{1}{2}\left[ 1+rq\,\,A(\widehat{a},\widehat{b},\lambda )\,B(\widehat{a},%
\widehat{b},\lambda )\right] 
\]
\begin{equation}
\qquad \qquad \qquad \quad =f_{DET}(r,q|\widehat{a},\widehat{b},\lambda )
\end{equation}
which is obtained by assuming that $E_{DET}^{(12)}(\widehat{a},\widehat{b}%
,\lambda )=A(\widehat{a},\widehat{b},\lambda )\,B(\widehat{a},\widehat{b}%
,\lambda )$, without introducing the parameter independence condition.
Since, the situation $q\ A(\widehat{a},\widehat{b},\lambda )=+1$ is only
allowed when $q=r\,$, the exchangeability condition cannot be established
here for $r\neq q$. Also, since the situation $r\ A(\widehat{a},\widehat{b}%
,\lambda )=-1$ never occures, the relation (21) reduces to the following
form,

\begin{equation}
p_{DET}^{(2)}(\sigma _{b}=q|\widehat{a},\widehat{b},\sigma _{a}=r,\lambda )=%
\frac{1}{2}\left[ 1+q\,\,B(\widehat{a},\lambda )\right]
=p_{DET}^{(2)}(\sigma _{b}=q|\widehat{a},\widehat{b},\lambda )
\end{equation}
which is equal to (5), if we introduce the assumption of determinism. Here,
the function $f_{DET}(r,q|\widehat{a},\widehat{b},\lambda )$ cannot
reproduce the conditional probability $P^{(2)}(\sigma _{b}=q|\widehat{a},%
\widehat{b},\sigma _{a}=r,\Psi _{0})$ in relation (15), and the
exchangeability condition at the quantum level does not contradict the
outcome independence condition at the hidden variables level. This shows
that \textit{the outcome independence condition is only feasible in a
deterministic framework,} a situation in which the exchangeability condition
is lost at the hidden variables level.\textit{\ }

\section{Appendix}

To prove (12), we introduce two dynamical variables $S_{1}$ and $S_{2}$
taking values $s_{1}$, $s_{2}=\pm 1$. One can express the joint
probabilities in terms of some mean values (denoted by $\overline{S_{1}}$, $%
\overline{S_{2}}$ and $\overline{S_{1}S_{2}}$),

\begin{equation}
\mathsf{P}(S_{1}=s_{1},S_{2}=s_{2})=\frac{1}{4}\left[ 1+s_{1}\ \overline{%
S_{1}}+s_{2}\ \overline{S_{2}}+s_{1}s_{2}\ \overline{S_{1}S_{2}}\right] 
\tag{A.1}
\end{equation}

The average quantities $\overline{S_{i}}$ ($i=1,2$) and $\overline{S_{1}S_{2}%
}$ were represented at the quantum level by $\langle S_{i}\rangle $ and $%
\langle S_{1}S_{2}\rangle $ and at the hidden variables level by $E^{(i)}$
and $E^{(12)}$, respectively. The marginal probability $\mathsf{P}%
(S_{1}=s_{1})$ is obtained by summing over $s_{2}$ in (A.1),

\begin{equation}
\mathsf{P}(S_{1}=s_{1})=\frac{1}{2}\left[ 1+s_{1}\ \overline{S_{1}}\right] 
\tag{A.2}
\end{equation}

Now, we can write (A.1) in the following form

\begin{equation}
\mathsf{P}(S_{1}=s_{1},S_{2}=s_{2})=\frac{1}{4}\left[ 1+s_{1}\overline{S_{1}}%
\right] \left[ 1+\frac{s_{2}\ \overline{S_{2}}+s_{1}s_{2}\ \overline{%
S_{1}S_{2}}}{1+s_{1}\ \overline{S_{1}}}\right]  \tag{A.3}
\end{equation}

Using Bayesian probability and comparing (A.3) with (A.2), we get the
conditional probability $\mathsf{P}(S_{2}=s_{2}|S_{1}=s_{1})$ as\bigskip

\begin{equation}
\mathsf{P}(S_{2}=s_{2}|S_{1}=s_{1})=\frac{1}{2}\left[ 1+\frac{s_{2}\ 
\overline{S_{2}}+s_{1}s_{2}\ \overline{S_{1}S_{2}}}{1+s_{1}\ \overline{S_{1}}%
}\right]  \tag{A.4}
\end{equation}

Exchanging $s_{1}$ and $s_{2}$ leads to the following relation\bigskip

\begin{equation}
\mathsf{P}(S_{2}=s_{1}|S_{1}=s_{2})=\frac{1}{2}\left[ 1+\frac{s_{1}\ 
\overline{S_{2}}+s_{1}s_{2}\ \overline{S_{1}S_{2}}}{1+s_{2}\ \overline{S_{1}}%
}\right]  \tag{A.5}
\end{equation}

Now if we impose the exchangeability condition (11) on (A.4) and (A.5), we
get (12) (considering that $s_{1}^{2}=s_{2}^{2}=1$), provided $\overline{%
S_{2}}=\overline{S_{1}}\ \overline{S_{1}S_{2}}.$

In an ideal EPRB thought experiment, the exchangeability condition (11)
holds because $\overline{S_{2}}=\overline{S_{1}}=0$ ($S_{1}=\sigma _{a},$ $%
S_{2}=\sigma _{b}$). There are some examples for which the relation $%
\overline{S_{2}}=\overline{S_{1}}\ \overline{S_{1}S_{2}}$ holds
nontrivially(see e.g. ref. [17]).

\end{document}